\documentclass[letterpaper]{article}
\usepackage{aaai}
\usepackage{amssymb}
\usepackage{epsfig}
\usepackage{times}
\usepackage{helvet}
\usepackage{courier}

\newtheorem{axiom}{Axiom}
\newtheorem{theorem}{Theorem}

\begin{document}
%
\title{Projective Expected Utility}
\author{Pierfrancesco La Mura\\
Department of Microeconomics and Information Systems\\
HHL - Leipzig Graduate School of Management\\
Jahnallee 59, 04109 Leipzig (Germany)\\
}
\maketitle
\begin{abstract}
\begin{quote}
Motivated by several classic decision-theoretic paradoxes, and by analogies with the paradoxes which in physics motivated the development of quantum mechanics, we introduce a projective generalization of expected utility along the lines of the quantum-mechanical generalization of probability theory. The resulting decision theory accommodates the dominant paradoxes, while retaining significant simplicity and tractability. In particular, every finite game within this larger class of preferences still has an equilibrium.
\end{quote}
\end{abstract}

\section{Introduction} John von Neumann (1903-1957) is widely regarded as a
founding father of many fields, including game theory, decision theory,
quantum mechanics and computer science. Two of his contributions are of
special importance in our context. In 1932, von Neumann proposed the first
rigorous foundation for quantum mechanics, based on a calculus of projections
in Hilbert spaces. In 1944, together with Oskar Morgenstern, he gave the first
axiomatic foundation for the expected utility hypothesis (Bernoulli 1738). In
both cases, the frameworks he pioneered are still at the core of the
respective fields. In particular, the expected utility hypothesis is still the
\emph{de facto} foundation of fields such as finance and game theory.

The von Neumann - Morgenstern axiomatization of expected utility, and later on
the subjective formulations by Savage (1954) and Anscombe and Aumann (1963)
were immediately greeted as simple and intuitively compelling. Yet, in the
course of time, a number of empirical violations and paradoxes (Allais 1953,
Ellsberg 1961, Rabin and Thaler 2001) came to cast doubt on the validity of
the hypothesis as a foundation for the theory of rational decisions in
conditions of risk and subjective uncertainty. In economics and in the social
sciences, the shortcomings of the expected utility hypothesis are generally
well known, but often tacitly accepted in view of the great tractability and
usefulness of the corresponding mathematical framework. In fact, the
hypothesis postulates that preferences can be represented by way of a utility
functional which is linear in probabilities, and linearity makes expected
utility representations particularly tractable in models and applications.

The experimental paradoxes, which in the context of physics motivated the
introduction of quantum mechanics, bear interesting relations with some of the
decision-theoretic paradoxes which came to challenge the status of expected
utility in the social sciences: in both cases, the anomalies can be regarded
as violations of an appropriately defined notion of independence. In physics,
quantum mechanics was introduced as a tractable and empirically accurate
mathematical framework in the presence of such violations. In economics and
the social sciences, the importance of accounting for violations of the
expected utility hypothesis has long been recognized, but so far none of its
numerous alternatives has emerged as dominant, sometimes due to a lack of
mathematical tractability, or to the ad-hoc nature of some axiomatic
proposals. Motivated by these considerations, we would like to introduce a
decision-theoretic framework which accommodates the dominant paradoxes while
retaining significant simplicity and tractability. As we shall see, this is
obtained by weakening the expected utility hypothesis to its projective
counterpart, in analogy with the quantum-mechanical generalization of
classical probability theory.

The structure of the paper is as follows. The next section briefly reviews the
von Neumann-Morgenstern framework. Sections 3 and 4, respectively, present
Allais' and Ellsberg's paradoxes. Section 5 introduces a mathematical
framework for projective expected utility, and a representation result.
Section 6 contains a subjective formulation for projective expected utility,
and a corresponding representation result. In section 7 there is a brief
discussion, and in sections 8 and 9, respectively, we show how Allais' and Ellsberg's paradoxes 
can be accommodated within the new framework. Section 10
discusses the multi-agent case and the issue of existence of strategic
equilibrium, while the last section concludes.

\section{von Neumann - Morgenstern Expected Utility} Let $S$ be a finite set
of outcomes, and $\Delta$ be the set of probability functions defined on $S,$
taken to represent risky prospects (or {\emph{lotteries}}). Next, let
$\succeq$ be a complete and transitive binary relation defined on $\Delta
\times \Delta,$ representing a decision-maker's preference ordering over
lotteries. Indifference of $p, q \in \Delta$ is defined as [$p \succeq q$ and
$q \succeq p$] and denoted as $p \sim q$, while strict preference of $p$ over
$q$ is defined as [$p \succeq q$ and not $q \succeq p$], and denoted by $p
\succ q$. The preference ordering is assumed to satisfy the following two
conditions.

\begin{axiom}
  (Archimedean) For all $p, q, r \in \Delta$ with $p \succ q \succ r,$ there
  exist $\alpha, \beta \in (0, 1)$ such that $\alpha p + (1 - \alpha) r \succ
  q \succ \beta p + (1 - \beta) r.$
\end{axiom}

\begin{axiom}
  (Independence) For all $p, q, r \in \Delta,$ $p \succeq q$ if, and only if,
  $\alpha p + (1 - \alpha) r \succeq \alpha q + (1 - \alpha) r$ for all
  $\alpha \in [0, 1] .$
\end{axiom}

{\bigskip}

A functional $u : \Delta \rightarrow R$ is said to represent $\succeq$ if,
for all $p, q \in \Delta,$ $p \succeq q$ if and only if $u (p) \geq u (q) .$

\begin{theorem}
  (von Neumann and Morgenstern) Axioms $1$ and $2$ are jointly equivalent to
  the existence of a functional $u : \Delta \rightarrow R$ which represents
  $\succeq$ such that, for all $p \in \Delta$,
  \[ u (p) = \sum_{s \in S} u (s) p (s) . \]
\end{theorem}

The von Neumann - Morgenstern setting is appropriate whenever the nature of
the uncertainty is purely objective: all lotteries are associated to objective
random devices, such as dice or roulette wheels, with well-defined and known
frequencies for all outcomes, and the decision-maker only evaluates a lottery
based on the frequencies of its outcomes.

\section{The ``Double-Slit Paradox'' of Decision Theory} The following paradox
is due to Allais (1953). First, please choose between:

\bigskip

A: A chance of winning $4000$ dollars with probability $0.2$ (expected value
$800$ dollars)

B: A chance of winning $3000$ dollars with probability $0.25$ (expected value
$750$ dollars).

{\bigskip}

Next, please choose between:

{\bigskip}

C: A chance of winning $4000$ dollars with probability $0.8$ (expected value
$3200$ dollars)

B: A chance of winning $3000$ dollars with certainty.

{\bigskip}

If you chose $A$ over $B,$ and $D$ over $C,$ then you are in the modal class
of respondents. The paradox lies in the observation that $A$ and $C$ are
special cases of a two-stage lottery $E$ which in the first stage either
returns zero dollars with probably $(1 - \alpha)$ or, with probability
$\alpha,$ it leads to a second stage where one gets $4000$ dollars with
probability $0.8$ and zero otherwise. In particular, if $\alpha$ is set to $1$
then $E$ reduces to $C,$ and if $\alpha$ is set to $0.25$ it reduces to $A.$
Similarly, $B$ and $D$ are special cases of a two-stage lottery $F$ which
again with probability $(1 - \alpha)$ returns zero, and with probability
$\alpha$ continues to a second stage where one wins $3000$ dollars with
probability $1.$ Again, if $\alpha = 1$ then $F$ reduces to $D,$ and if
$\alpha = 0.25$ it reduces to $B.$ Then it is easy to see that the $[A \succ
B,$ $D \succ C]$ pattern violates Axiom 2 (Independence), as $E$ can be
regarded as a lottery $\alpha p + (1 - \alpha) r$, and $F$ as a lottery
$\alpha q + (1 - \alpha) r$, where $p$ and $q$ represent lottery $C$ and $D$
respectively and $r$ represents the lottery in which one gets zero dollars
with certainty. When comparing $E$ and $F$, why should it matter what is the
value of $\alpha ?$ Yet, experimentally one finds that it does.

{\bigskip}

Allais' paradox bears a certain resemblance with the well-known double-slit
paradox in physics. Imagine cutting two parallel slits in a sheet of
cardboard, shining light through them, and observing the resulting pattern as
particles going through the two slits scatter on a wall behind the cardboard
barrier. Experimentally, one finds that when both slits are open, the overall
scattering pattern is not the sum of the two scattering patterns produced when
one slit is open and the other closed. The effect does not go away even if
particles of light are shone one by one, and this is paradoxical: why should
it matter to an individual particle which happens to go through the left slit,
when determining where to scatter later on, with what probability it could
have gone through the right slit instead? In a sense, each particle in the
double-slit experiment behaves like a decision-maker who violates the
Independence axiom in Allais' experiment.

\section{Ellsberg's Paradox} Another disturbing violation of the expected
utility hypothesis was pointed out by Ellsberg (1961). Suppose that a box
contains $300$ balls of three possible colors: red ($R$), green ($G$), and
blue ($B$). You know that the box contains exactly $100$ red balls, but are
given no information on the proportion of green and blue.

You win if you guess which color will be drawn. Do you prefer to bet on red
($R$) or on green ($G$)? Many respondents choose $R$, on grounds that the
probability of drawing a red ball is known to be $1 / 3$, while the only
information on the probability of drawing a green ball is that it is between
$0$ and $2 / 3.$ Now suppose that you win if you guess which color will
\emph{not} be drawn. Do you prefer to bet that red will not be drawn
($\overline{R}$) or that green will not be drawn ($\overline{G}$)? Again many
respondents prefer to bet on $\overline{R}$, as the probability is known ($2 /
3$) while the probability of $\overline{G}$ is only known to be between $1$
and $1 / 3.$

The pattern [$R \succ G$, $\overline{R} \succ \overline{G}$] is incompatible
with von Neumann - Morgenstern expected utility, which only deals with known
probabilities, and is also incompatible with the Savage (1954) formulation of
expected utility with subjective probability as it violates its Sure Thing
axiom. The paradox suggests that, in order to account for such patterns of
choice, subjective uncertainty and risk should be handled as distinct notions.
Once again, the issue bears interesting relations with quantum mechanics,
where the need to keep a distinction between subject-related uncertainty and
objective risk also presented itself, and is naturally met in the formalism by
the distinction between pure and mixed states.

\section{Projective Expected Utility} Let $X$ be the positive orthant of the
unit sphere in $\mathbb{R}^n$, where $n$ is the cardinality of the set of
relevant outcomes $S : = \left\{ s_1, s_2, ..., s_n \right\} .$ Then von
Neumann - Morgenstern lotteries, regarded as elements of the unit simplex, are
in one-to-one correspondence with elements of $X,$ which can therefore be
interpreted as risky prospects, for which the frequencies of the relevant
outcomes are fully known. Observe that, while the projections of elements of
the unit simplex (and hence, $L^1$ unit vectors) on the basis vectors can be
naturally associated with probabilities, if we choose to model von Neumann -
Morgenstern lotteries as elements of the unit sphere (and hence, as unit
vectors in $L^2$) then probabilities are naturally associated with
\emph{squared} projections. The advantage of such move is that $L^2$ is
the only $L^p$ space which is also a Hilbert space, and Hilbert spaces have a
very tractable projective structure which is exploited by the representation.
In particular, it is unique to $L^2$ that the set of unit vectors is invariant
with respect to projections.

Next, let $\left\langle .|. \right\rangle$ denote the usual inner product in
$\mathbb{R}^n$. An orthonormal basis is a set of unit vectors $(b_1, ...,
b_n)$ such that $\left\langle b_i |b_j \right\rangle = 0$ whenever $i \neq j$.
The natural basis corresponds to the set of degenerate lotteries returning
each objective lottery outcome with certainty, and is conveniently identified
with the set of objective lottery outcomes. Yet, in any realistic experimental
setting, it is very unlikely that those objective outcomes will happen to
coincide with the relevant outcomes as subjectively perceived by the
decision-maker. Moreover, even if those could be fully elicited, it would be
generally problematic to relate a von Neumann-Morgenstern lottery, which only
specifies the probabilities of the objective outcomes, with the
decision-relevant probabilities induced on the subjective outcomes. The
perspective of the observer or modeller is inesorably bound to objectively
measurable entities, such as frequencies and prizes; by contrast, the
decision-maker thinks and acts based on subjective preferences and subjective
outcomes, which in a revealed-preference context should be presumed to exist
while at the same time assumed, as a methodological principle, to be
unaccessible to direct measurement.

\begin{axiom}
  (Born's Rule) There exists an orthonormal basis $(z_1, ..., z_n)$ such that,
  for all $x \in X$ and all $s_i \in S,$ any two lotteries are indifferent
  whenever their risk profiles
  \[ p_x (z_i) = \left\langle x|z_i \right\rangle^2, i = 1, \ldots, n \]
  coincide.
\end{axiom}

Axiom 3 requires that there exist an interpretation of elements of $X$ in
terms of the relevant dimensions of risk as perceived by the decision-maker.
In the von Neumann - Morgenstern treatment, Axiom 3 is tacitly assumed to hold
with respect to the natural basis in $\mathbb{R}^n .$ This implicit assumption
amounts to the requirement that lotteries are only evaluated based on the
probabilities they induce on the objective lottery outcomes.

By contrast, the \emph{preferred basis} postulated in Axiom 3 is allowed
to vary across different decision-makers, capturing the idea that the relevant
dimensions of risk (that is, those pertaining to the actual subjective
consequences) may be perceived differently by different subjects, perhaps due
to portfolio effects, while the natural basis may be taken to represent the
relevant dimensions of risk as perceived by the modeler or an external
observer (that is, in terms of the objective outcomes as evaluated by the
observer). Portfolio effects are very difficult to exclude in an experimental
setting. For instance, the very fact of proposing to a subject the Allais or
Ellsberg games described above generates an expectation of gain. An invitation
to play the game can be effectively regarded as a risky security which is
donated to the subject, and whose subjective returns are obviously correlated,
but do not necessarily coincide, with the monetary outcomes of the experiment.
Even in such simple contexts, significant hedging behavior cannot be in
principle excluded.

In our context, orthogonality captures the idea that two events or outcomes
are mutually exclusive (for one event to have probability one, the other must
have probability zero). The preferred basis captures which, among all possible
ways to partition the relevant uncertainty into a set of mutually exclusive
events or outcomes, leads to a set of payoff-relevant orthogonal lotteries
from which preferences on all other lotteries can be assigned in a linear
fashion. In particular, the basis elements must span the whole range of
preferences. For instance, in case a decision-maker is indifferent between two
outcomes, but strictly prefers to receive both of them with equal frequency,
preferences on the two outcomes do not span all the relevant range; therefore,
this preference pattern cannot be captured in the natural basis, and hence in
von Neumann - Morgenstern expected utility. Axiom 3 postulates that each
lottery is evaluated by the decision maker solely by the uncertainty it
induces on appropriately chosen payoff-relevant dimensions. For now, and only
for simplicity, we assume that the cardinality of the set of subjective
outcomes coincides with that of the objective ones; we relax this assumption
later on, in the subjective formulation.

Once an orthonormal basis Z is given, each objective lottery $x$ uniquely
corresponds to a function $p_x : Z \rightarrow [0, 1],$ such that $p_x (z_i) =
\left\langle x|z_i \right\rangle^2$ for all $z_i \in Z.$ Let $B$ be the set of
all such {\emph{risk profiles}} $p_x,$ for $x \in X,$ and let $\succeq$ be
the complete and transitive preference ordering induced on $B \times B$ by
preferences on the underlying lotteries. We postulate the following two
axioms, which mirror those in the von Neumann - Morgenstern treatment.

\begin{axiom}
  (Archimedean) For all $x, y, z \in X$ with $p (x) \succ p (y) \succ p (z),$
  there exist $\alpha, \beta \in (0, 1)$ such that $\alpha p (x) + (1 -
  \alpha) p (z) \succ p (y) \succ$ $\beta p (x) + (1 - \beta) p (z) .$
\end{axiom}

{\bigskip}

\begin{axiom}
  (Independence) For all $x, y, z \in X$, $p_x \succeq p_y$ if, and only if,
  $\alpha p_x + (1 - \alpha) p_z \succeq ap_y + (1 - \alpha) p_z$ for all
  $\alpha \in [0, 1] .$
\end{axiom}

Observe that the two axioms impose conditions solely on risk profiles, and not
on the underlying lotteries. In particular, note that a convex combination
$\alpha p_x + (1 - \alpha) p_y$, where $p_x$ and $p_y$ are objective
probability functions, is also an objective probability function. Hence, the
type of mixing in Axioms 4 and 5 can be regarded as objective, while
subjective mixing will be later on captured by convex combinations (in $R^n$)
of the underlying (Anscombe-Aumann) acts.

\begin{theorem}
  Axioms 3-5 are jointly equivalent to the existence of a symmetric matrix $U$
  such that $u (x) : = x' Ux$ for all $x \in X$ represents $\succeq$.
\end{theorem}

\textbf{Proof.}
  Assume that Axiom 3 holds with respect to a given orthonormal basis $(z_1,,
  z_n)$. By the von Neumann - Morgenstern result (which applies to any convex
  mixture set, such as $B$) Axioms 4 and 5 are jointly equivalent to the
  existence of a functional $u$ which represents the ordering and is linear in
  $p$, i.e.
  \[ u (x) = \sum_{i = 1}^n u (s_i) p_{s_i} (x) = \sum_{i = 1}^n u (s_i)
     \left\langle x|z_i \right\rangle^2, \]
  where the second equality is by definition of $p$ as squared inner product
  with respect to the preferred basis. The above can be equivalently written,
  in matrix form, as
  \[ u (x) = x' P' DPx = x' Ux, \]
  where $D$ is the diagonal matrix with the payoffs on the main diagonal, $P$
  is the projection matrix associated to $(z_1,, z_n)$, and $U : = P' DP$ is
  symmetric. Conversely, by the Spectral Decomposition theorem, for any
  symmetric matrix $U$ there exist a diagonal matrix $D$ and a projection
  matrix $P$ such that $U = P' DP$, and hence
  \[ x' Ux = x' P' DPx \]
  for all $x \in X.$ But this is just expected utility with respect to the
  orthonormal basis defined by $P$. Hence, the three axioms are jointly
  equivalent to the existence of a symmetric matrix $U$ such that $u (x) : =
  x' Ux$ represents the preference ordering. QED

\section{Subjective Formulation} We introduce the following setup and
notation.

$\mathcal{S}$ is a finite set of states of Nature.

$\left. \langle .|. \right\rangle$ denotes the usual inner product in
Euclidean space.

$\Omega$ is the natural basis in $\mathbb{R}^n$, identified with a finite set
$\{\omega^1, \ldots ., \omega^n \}$ of lottery outcomes (prizes).

$Z$ is an orthonormal basis in $\mathbb{R}^m$, with $m \geq n,$ identified
with a finite set of subjective outcomes or consequences $\{z^1, \ldots ., z^m
\}$. $V$ is an arbitrary $(m \times n)$ matrix chosen so that, for all
$\omega^i$ in $\Omega,$ $V \omega^i $ is a unit vector in $\mathbb{R}^m$.

The quantity $\left\langle z^j |V \omega^i_{^{}} \right\rangle^2$ is
interpreted as $p (z^j | \omega^i)$, the conditional probability of subjective
outcome $z^j$ given objective outcome $\omega^i_{^{}}$.

Lotteries correspond to $L^2$ unit vectors $x \in \mathbb{R}^n_+$; $X$ is the
set of all lotteries.

Since $\Omega$ is the natural basis, $\left. \left. \right\langle \omega^i |x
\right\rangle^2 = x^2_i$; this quantity is interpreted as $p (\omega^i |x)$.

\bigskip

Once the subjective consequences $z^j$ are specified, for any lottery $x$ one
can readily compute $p (\omega^i |x) = x_i^2$ and $p (z^j |x) = \sum_i p
(\omega^i |x) p (z^j | \omega^i) = \sum_i x_i^2 \left\langle z^j |V
\omega^i_{^{}} \right\rangle^2$. Moreover, given the latter probabilistic
constraints, one can readily identify the unique lottery $x$ and an
orthonormal basis $Z$ which jointly satisfy them. Hence, in the above
construction lotteries are identified with respect to two different frames of
reference: objective lottery outcomes, and subjective consequences.

\bigskip

An act is identified with a function $f : \mathcal{S} \rightarrow X$. $H$ is the set of all acts.

$\Delta (X)$ is the (nonempty, closed and convex) set of all probability
functions on $Z$ induced by lotteries in $X$.

$M$ is the set of all vectors $(p_s)_{s \in S}$, with $p_s \in \Delta (X)$.

\bigskip

For each $f \in H$ a corresponding {\emph{risk profile}} $p^f \in M$ is
defined, for all $s \in \mathcal{S}$ and all $z^j \in Z,$ by $p^f_s
(z^j):= \left\langle z^j |V f_s \rangle^2 \right.$. Observe that, even
though $p_s^f$ is a well-defined probability function, $p^f_s (z^j)$ generally
differs from the probability of $z^j$ given $f (s)$, which is given by $\sum_i
f_i (s)^2 \left\langle z^j |V \omega^i_{^{}} \right\rangle^2$. Hence, except
when the two measures happen to coincide, the risk profile $p^f$ cannot be
interpreted as a vector of objective probabilities, but rather as a possible
sufficient statistic for the ranking of acts from the point of view of the
decision-maker.

\bigskip

As customary, we assume that the decision-makers preferences are
characterized by a rational ({\emph{i.e.}}, complete and transitive)
preference ordering $\succeq$ on acts.

Next, we proceed with the following assumptions, which mirror those in
Anscombe and Aumann (1963).

\begin{axiom}
  (Projective) There exists an orthonormal basis $Z$ such that any two acts
  $f, g \in H$ are indifferent if $p^f = p^g .$
\end{axiom}

In Anscombe and Aumann's setting, the above axiom is implicitly assumed to
hold with $Z \equiv \Omega$ and $V \equiv I$, where $I$ is the $(n \times n)$
identity matrix. Because of Axiom 6, preferences on acts can be equivalently
expressed as preferences on risk profiles. For all $p^f, p^g \in M$, we
stipulate that $p^f \succeq p^g$ if and only if $f \succeq g$.

\begin{axiom}
(Archimedean) If $p^f, p^g, p^h \in M$ are such that $p^f \succ p^g
  \succ p^h$, then there exist $a, b \in (0, 1)$ such that $a p^f + (1 - a)
  p^h \succ p^g \succ b p^f + (1 - b) p^h$.
\end{axiom}

\begin{axiom}
  (Independence) For all $p^f, p^g, p^h \in M$, and for all $a \in (0, 1]$,
  $p^f \succ p^g$ if and only if $a p^f + (1 - a) p^h \succ a p^g + (1 - a)
  p^h$.
\end{axiom}

\begin{axiom}
  (Non-degeneracy) There exist $p^f, p^g \in M$ such that $p^f \succ p^g$.
\end{axiom}

\begin{axiom}
  (State independence) Let $s, t \in \mathcal{S}$ be non-null states, and let
  $p, q \in \Delta (X)$. Then, for any $p^f \in M$,

\smallskip
 
$(p^f_1,..., p^f_{s
  - 1}, p, p^f_{s + 1},..., p^f_n) \succ (p^f_1,..., p^f_{s
  - 1}, q, p^f_{s + 1},..., p^f_n)$ 

\smallskip

if, and only if,

\smallskip

$(p^f_1,...,
  p^f_{t - 1}, p, p^f_{t + 1},..., p^f_n) \succ (p^f_1,...,
  p^f_{t - 1}, q, p^f_{t + 1},..., p^f_n)$.
\end{axiom}

\begin{theorem}
  (Anscombe and Aumann) The preference relation $\succeq_{}$ fulfills Axioms
  $6 - 10$ if and only if there is a unique probability measure $\pi$ on
  $\mathcal{S}$ and a non-constant function $u : X \rightarrow R$ (unique up
  to positive affine rescaling) such that, for any $f, g \in H,$ $f \succeq
  g$ if and only if
  
  $\sum_{s \in \mathcal{S}} \pi (s) \sum_{z^i \in Z} p^f_s (z^i) u (z^i) \geq
  \sum_{s \in \mathcal{S}} \pi (s) \sum_{z^i \in Z} p^g_s (z^i)^{} u (z^i) .$
\end{theorem}

\begin{theorem}
  The preference relation $\succeq_{}$ fulfills Axioms $6 - 10$ if and only
  if there is a unique probability measure $\pi$ on $\mathcal{S}$ and a
  symmetric $(n \times n)$ matrix $U$ with non-constant eigenvalues such that,
  for any $f, g \in H,$ $f \succeq g$ if and only if $\sum_{s \in
  \mathcal{S}} \pi (s) f_s' U f_s \geq \sum_{s \in \mathcal{S}} \pi (s) g_s' U
  g_s .$
\end{theorem}

\textbf{Proof}.
  Let D be the $(m \times m)$ diagonal matrix defined by $D_{i, i} = u (z^i),$
  and let $P$ be the projection matrix defined by $(P_{i, .})' = z^i .$ If
  Axioms 6-10 hold, we know from Theorem 3 that the preference ordering has an
  expected utility representation. Observe that, since $(P V f_s)_i^2 = \left.
  \right\langle z^i |V f_s \rangle^2 = p^f_{_s} (z^i),$ $\sum_{s \in
  \mathcal{S}} \pi (s) \sum_{z^i \in Z} p^f_s (z^i) u (z^i) = \sum_{s \in
  \mathcal{S}} \pi (s) f_s' V' P' D P V f_s .$
  
  It follows that the expected utility of any act $f$ can be written as
  $\sum_{s \in \mathcal{S}} \pi (s) f_s' U f_s,$ where $U : = V' P' D P V$ is
  a $(n \times n)$ symmetric matrix. Since the diagonal elements of $D$ are
  the eigenvalues of $U,$ the latter matrix has non-constant eigenvalues.
  
  Conversely, let $U$ be a symmetric $(n \times n)$ matrix with non-constant
  eigenvalues. By the Spectral Decomposition theorem there exist a projection
  matrix $P$ and a diagonal matrix $D$ such that $U = V' P' D P V,$ and
  therefore $\sum_{s \in \mathcal{S}} \pi (s) f_s' U f_s$=$\sum_{s \in
  \mathcal{S}} \pi (s) f_s' V' P' D P V f_s .$ Observe that the right-hand
  side is the expected utility of $f$ with respect to the orthonormal basis
  $Z$ defined by $z^i  = P_{i, .}',$and with $u (z^i)= D_{i, i} .$
  Since the diagonal elements of $D$ are the eigenvalues of $U,$ if the latter
  has non-constant eigenvalues then $u (z^i)$ is also non-constant, and hence
  by Theorem 3 Axioms 6-10 must hold. QED

\bigskip

The above formulation extends the representation to situations of subjective
uncertainty. Say that an act is $pure$ or $certain$ if it returns for sure a
given objective lottery, and $mixed$ or $uncertain$ if it can be obtained as
the convex combination $\alpha f + (1 - \alpha) g$ of two non-identical acts
$f$ and $g.$

\section{Properties of the Representation} Our representation generalizes the
Anscombe-Aumann expected utility framework in three directions. First,
subjective uncertainty (from mixed acts) and risk (from pure acts) are treated
as distinct notions. Second, as we shall see, within this class of preferences
both Allais' and Ellsberg's paradoxes are accommodated. Finally, the
construction easily extends to the complex unit sphere, provided that
$\left\langle x|y \right\rangle^2$ is replaced by $| \left\langle x|y
\right\rangle |^2$ in the definition of $p,$ in which case Theorems 2 and 4
hold with respect to a Hermitian (rather than symmetric) payoff matrix $U$,
and the result also provides axiomatic foundations for decisions involving
quantum uncertainty.

\bigskip

Let $e_i,$ $e_j$ be the degenerate lotteries assigning probability $1$ to
outcomes $s_i$ and $s_j,$ respectively, and let $e_{i, j}$ be the lottery
assigning probability $1 / 2$ to each of the two states. Observe that, for any
two distinct $s_i$ and $s_j,$
\[ U_{ij} = u (e_{i, j}) - ( \frac{1}{2} u (e_i) + \frac{1}{2} u (e_j)) . \]
It follows that the off-diagonal entry $U_{ij}$ in the payoff matrix can be
interpreted as the discount, or premium, attached to an equiprobable
combination of the two outcomes with respect to its expected utility
base-line, and hence as a measure of preference for risk versus uncertainty
along the specific dimension involving outcomes $s_i$ and $s_j .$

\bigskip

Observe that the functional in Theorem 2 is quadratic in $x,$ but linear in
$p.$ If $U$ is diagonal, then its eigenvalues coincide with the diagonal
elements. In von Neumann - Morgenstern expected utility, those eigenvalues
contain all the relevant information about the decision-maker's risk
attitudes. In our framework, risk attitudes are jointly captured by both the
diagonal and non-diagonal elements of $U$, which are completely characterized
by a diagonal matrix $D$ with the eigenvalues of $U$ on the main diagonal and
a projection matrix $P.$ Furthermore, in our setting attitudes towards
uncertainty are captured by the concavity or convexity (in $x$) of the
quadratic form $x' Ux,$ and therefore, ultimately, by the definiteness
condition of $U$ and the sign of its eigenvalues. Convexity corresponds to the
case of all positive eigenvalues, and captures the idea that risk is
\emph{ceteris paribus} preferred to uncertainty, while concavity
corresponds to the case of all negative eigenvalues and captures the opposite
idea.

\section{Example: Objective Uncertainty}

Figure 1 below presents several examples of indifference maps on pure
lotteries which can be obtained within our class of preferences for different
choices of $U$.

\begin{figure}[h]
  \epsfig{file=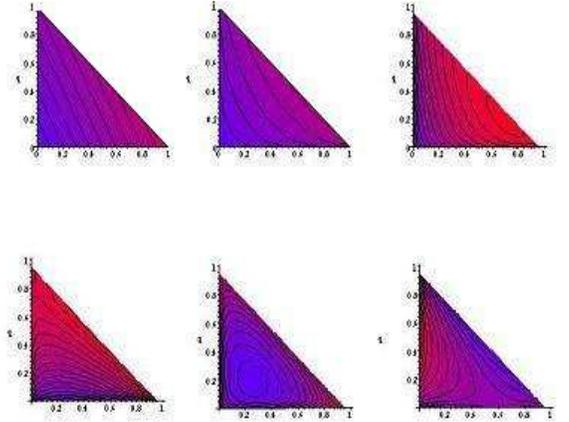,height=8cm,width=10cm}
  \caption{Examples of indifference maps on the probability triangle}
\end{figure}

The first pattern (parallel straight lines) characterizes von Neumann -
Morgenstern expected utility. Within our class of representations, it
corresponds to the special case of a diagonal payoff matrix $U.$ All other
patterns are impossible within von Neumann - Morgenstern expected utility.

\bigskip

The representation is sufficiently general to accommodate Allais' paradox from
section 3. In the context of the example in section 3, let $\{s_1, s_2, s_3
\}$ be the states in which $4000,$ $3000,$ and $0$ dollars are won,
respectively. To accommodate Allais' paradox, assume a slight aversion to the
risk of obtaining no gain ($s_3$):

\bigskip

$U = \begin{array}{c}
  \\
  s_1\\
  s_2\\
  s_3
\end{array} \begin{array}{c}
  s_1\\
  13\\
  0\\
  - 1
\end{array} \begin{array}{c}
  s_2\\
  0\\
  10\\
  - 1
\end{array} \begin{array}{c}
  s_3\\
  - 1\\
  - 1\\
  0
\end{array}$

\bigskip

Let the four lotteries $A,$ $B,$ $C,$ $D$ be defined, respectively, as the
following unit vectors in $S_+$: $a = ( \sqrt{0.2}, 0, \sqrt{0.8})$; $b = (0,
\sqrt{0.25}, \sqrt{0.75})$; $c = ( \sqrt{0.8}, 0, \sqrt{0.2})$; $d = (0, 1,
0)$. Then lottery $A$ is preferred to $B,$ while $D$ is preferred to $C$, as

\bigskip

$u (a) = a' Ua = 1.8,$

$u (b) = b' Ub = 1.634,$

$u (c) = c' Uc = 9.6,$

$u (d) = d' Ud = 10.$

\section{Example: Subjective Uncertainty} In the Ellsberg puzzle, suppose that
either all the non-red balls are green $(R = 100, G = 200, B = 0),$ or they
are all blue $(R = 100, G = 0, B = 200),$ with equal subjective probability.
Further, suppose that there are just two payoff-relevant outcomes, Win and
Lose. Then the following specification of the payoff matrix accommodates the
paradox.

\bigskip

$U = \begin{array}{c}
  \\
  \ $Win$\\
  \ $Lose$
\end{array} \begin{array}{c}
  \ $Win$\\
  1\\
  \alpha
\end{array} \begin{array}{c}
  \ $Lose$\\
  \alpha\\
  0
\end{array}$

\bigskip

As we shall see, if $\alpha = 0$ we are in the expected utility case, where
the decision-maker is indifferent between risk and uncertainty; when $\alpha >
0,$ risk is preferred to uncertainty; and when $\alpha < 0$, the
decision-maker prefers uncertainty to risk.

In fact, let $\left\{ Urn1, Urn2 \right\}$ be the set of
possible states of nature, with uniform subjective probability, and let 

\bigskip

$r = ( \sqrt{1 / 3}, \sqrt{2 / 3}),$

$\overline{r} = ( \sqrt{2 / 3}, \sqrt{1 / 3})$

\bigskip

be the lotteries representing $R$ and $\overline{R},$ respectively.
Furthermore, let

\bigskip

$w = \left( 1, 0), \right.$

$l = (0, 1)$

\bigskip

be the lotteries corresponding to a sure win and a sure loss, respectively.

\bigskip

The mixed acts $G$ and $\overline{G} $ have projective expected utilities
given by

\bigskip

$u (G) = p (Urn1) u ( \overline{r})$+$p (Urn2) u (l)$,

$u ( \bar{G}) = p (Urn1) u (r)$+$p (Urn2) u (w)$.

\bigskip

One also has that

\bigskip

$u (w) = 1,$

$u (l) = 0$,

$u (r) = r' Ur = 1 / 3 + \alpha \sqrt{8} / 3,$

$u ( \overline{r}) = \overline{r}' U \overline{r} = 2 / 3 + \alpha \sqrt{8} /
3,$

\bigskip

and therefore

\bigskip

$u (g) = 1 / 3 + \alpha \sqrt{2} / 3$

$u ( \overline{g}) = 2 / 3 + \alpha \sqrt{2} / 3.$

\bigskip

It follows that, whenever $\alpha > 0,$ $R$ is preferred to $G$ and
$\overline{R}$ to $\bar{G}$, so the paradox is accommodated. When $\alpha <
0,$the opposite pattern emerges: $G$ is preferred to $R$ and $\bar{G}$ to
$\overline{R}$. Finally, when $\alpha = 0$ the decision-maker is indifferent
between $R$ and $G,$ and between $\overline{R}$ and $\bar{G}$.

\section{Multi-Agent Decisions and Equilibrium} Within the class of
preferences characterized by Theorem 4, is it still true that every finite
game has a Nash equilibrium? If the payoff matrix $U$ is diagonal we are in
the classical case, so we know that any finite game has an equilibrium, which
moreover only involves objective risk (in our terms, this type of equilibrium
should be referred to as ``pure'', as it involves no subjective uncertainty).
For the general case, consider that $u (f)$ is still continuous and linear
with respect to the subjective beliefs $\pi$, while possibly nonlinear with
respect to risk. Therefore, all the necessary steps in Nash's proof (showing
that the best response correspondence is non-empty, convex-valued and
upper-hemicontinuous in order to apply Kakutani's fixed-point theorem) also
follow in our case. Hence, any finite game has an equilibrium even within this
larger class of preferences, although the equilibrium may not be pure (in our
sense): in general, an equilibrium will rest on a combination of objective
randomization and subjective uncertainty about other players' decisions.

\section{Conclusions} We presented a projective generalization of expected
utility, and showed that it can accommodate the dominant decision-theoretic
paradoxes. Whereas other generalizations of expected utility are typically
non-linear in probabilities, projective expected utility is possibly nonlinear
with respect to risk, but linear with respect to subject-related uncertainty.
We found that within this class of preferences the dominant paradoxes can be
accommodated, and every finite game still has an equilibrium. Moreover, the
projective calculus associated with the representation should make it
generally quite tractable in applications.

Our generalization of expected utility is closest to the one in Gyntelberg
and Hansen (2004), which is obtained in a Savage context by postulating a
non-classical (that is, non-Boolean) structure for the relevant events. By
contrast, our representation is obtained in an Anscombe-Aumann context, and
does not impose specific requirements on the nature of the relevant
uncertainty. In particular, in our context the dominant paradoxes can be
resolved even if the relevant uncertainty is of completely classical nature.
Furthermore, in case the event space is non-classical, our result also
provides foundations for decisions involving quantum uncertainty.

\section{References}

Allais, M. 1953. Le Comportement de l'Homme Rationnel devant le
Risque: Critique des postulats et axiomes de l'\'Ecole Americaine.
{\it Econometrica} 21: 503-546.

Bernoulli, D. 1738. Specimen theoriae novae de mensura sortis. In
Commentarii Academiae Scientiarum Imperialis Petropolitanae,
5: 175-192. Translated as ``Exposition of a new theory of the
measurement of risk''. 1954, {\it Econometrica} 22: 123-136.

Ellsberg, D. 1961. Risk, Ambiguity and the Savage Axioms.
{\it Quarterly Journal of Economics} 75: 643-669.

Gyntelberg, J. and Hansen, F. 2004. Expected utility theory with small
worlds,Department of Economics Discussion Paper 04-20,  University of Copenhagen.
http://ideas.repec.org/p/kud/kuiedp/0420.html

von Neumann, J. 1932. {\it Mathematische Grundlagen der
Quantenmechanic}. Berlin, Springer. Translated as {\it Mathematical
foundations of quantum mechanics}, 1955, Princeton University Press, Princeton
NJ.

von Neumann, J., and Morgenstern, O. 1944. {\it Theory of Games and
Economic Behavior}. Princeton University Press, Princeton NJ.

Rabin, M. and Thaler, R. H. 2001. Anomalies: Risk Aversion.
{\it Journal of Economic Perspectives} 15: 219-232.

Savage, L. J. 1954. {\it The Foundations of Statistics}. Wiley, New
York, NY.

\end{document}